\documentclass[aps,prb,a4paper,floatfix,showkeys,reprint,superscriptaddress]{revtex4-1}
\usepackage{times}
\usepackage{amssymb,amsmath,mathrsfs}
\usepackage{natmove}
\usepackage{natbib}
\usepackage[pdftex]{graphicx}
\usepackage{bm}
\usepackage[bookmarks, 
plainpages=false]{hyperref}
\usepackage{graphicx}
\usepackage[usenames,dvipsnames]{color}
\usepackage{epstopdf}
\usepackage{soul}
\usepackage{cleveref}
\usepackage[nolist,nohyperlinks]{acronym}
\usepackage[normalem]{ulem}
\crefname{figure}{Fig.}{Figs.}
\newcommand{\Gsum}{\tensor{\mathcal{G}}}
\newcommand{\Gneq}{\tensor{\mathcal{G}}^{\neq}}

\newcommand{\sub}[1]{_{\mathrm{#1}}}
\newcommand{\unit}[1]{\,\,{\mathrm{#1}}}

\newcommand{\alphat}{\tensor{\alpha}}
\DeclareMathAlphabet{\mathpzc}{OT1}{pzc}{m}{it}

\begin{document}
\title{Optical properties of 2D magnetoelectric point scattering lattices}
\author{Per Lunnemann}
\affiliation{DTU Fotonik, Department of Photonics Engineering, \O stedsplads 343, DK-2800, Denmark}
\author{Ivana Sersic}
\affiliation{Center for Nanophotonics, FOM Institute AMOLF,
Science Park 104, 1098 XG Amsterdam, The Netherlands}
\author{A. Femius Koenderink}
\email{fkoenderink@amolf.nl}
\affiliation{Center for Nanophotonics, FOM Institute AMOLF,
Science Park 104, 1098 XG Amsterdam, The Netherlands}
\date{\today}

\begin{abstract}
We explore the electrodynamic coupling between a plane wave and an infinite two-dimensional periodic lattice of magneto-electric point scatterers, deriving a semi-analytical theory with consistent treatment of  radiation damping, retardation, and energy conservation. We apply the theory to arrays of split ring resonators and provide a quantitive comparison of measured and calculated transmission spectra  at normal incidence as a function of lattice density, showing excellent agreement. We further show angle-dependent  transmission calculations for circularly polarized light  and compare with the angle-dependent response of a single split ring resonator, revealing the importance of  cross coupling between electric dipoles and magnetic dipoles for quantifying the pseudochiral response under oblique incidence of split ring lattices.
\end{abstract}
\date\today

\maketitle

\section{Introduction}
Since the seminal works of \textcite{Veselago1968} and Pendry\cite{Pendry2000}, much effort has been put into designing and fabricating artificial materials using periodic nanostructured materials with effective material parameters  $\bm{\epsilon}$  and  $\bm{\mu}$ that otherwise do not exist in nature\cite{Zheludev2012a}.  The mathematical tools  of transformation optics~\cite{Pendry2012} state that  full control over $\bm{\epsilon}$ and $\bm{\mu}$  allows nearly  arbitrarily rerouting of light through space\cite{Pendry2012}, with exotic applications such as superlenses and cloaking. Besides tailoring of $\bm{\epsilon}$ and $\bm{\mu}$, the scattering properties of the sub-wavelength building blocks that were developed for metamaterials have attracted much attention~\cite{Liu2008,Liu2009,Plum2009,Feth2010,Langguth2011,Powell2011,Decker2009,Decker2011,Cube2013}. Tailoring of the optical scattering properties  may be achieved by structural design of the scatterers to control their electric and magnetic dipole polarizability, as well as by tuning their mutual optical coupling by changing their relative coordination and orientation. With recent advances in nanotechnological fabrication techniques, based on these principles, novel metasurfaces\cite{Kildishev2013,Yu2012a} have been demonstrated, as well as  compact and on-chip compatible optical antennas\cite{Koenderink2009,Arango2012}, waveguides\cite{Maier2005}, flat lenses\cite{Yu2011,Aieta2012} and materials with  giant birefringence.\cite{Gansel2009,Yu2012a,Plum2009,Kats2012,Liu2009,Schaferling2012,Plum2011}

Scattering experiments on metamaterials are frequently done on periodic planar arrays of  scatterers with sub-diffraction pitch\cite{Lahiri2010,Klein2006,Linden2004,Enkrich2005,Sersic2009}. The chain of reasoning from measurement to effective media parameters generally starts from measured intensity reflection and transmission that are used to validate brute force \ac{FDTD} simulations.\cite{Capolino2009,Zhao2010,Smith2005} The \ac{FDTD} calculations in turn lead to retrieval of effective parameters from the calculated amplitude reflection and transmission.\cite{Linden2004,Enkrich2005} 
In vein of the classical Lorentz oscillator model, it is desirable to express array response in terms of a polarizability per element, rather than in an effective  $\bm{\epsilon}$ and   $\bm{\mu}$. Indeed,  it is now generally accepted that the \ac{SRR}  for instance is a strongly polarizable  electric and magnetic dipole scatterer, and that SRRs  interact depending on their density, local lattice coordination and relative orientation via near and far field dipole terms\cite{Liu2009,Sersic2009,Feth2010,Linden2004,Rockstuhl2006,Feth2010,Decker2009,Liu2008}.
Since split rings have  extinction cross sections far in excess of the typical unit cell areas of the metamaterial lattices they are stacked in, and comparable to the unitary limit\cite{Husnik2008,Sersic2009}, coupling is not only via $1/r^3$ near field interactions,\cite{Novotny2008} but also  via retarded far field terms.\cite{Sersic2009,Decker2011,Sersic2011} Indeed, transmission experiments on \acp{SRR} show strong superradiant broadening effects that increase with SRR density,\cite{Sersic2009,Feth2010} and further depend on incidence angle.  \textcite{Decker2011} attempted to account for these interactions using numerical summation of retarded electric dipole-dipole interactions on a 1D chain. However, in this approach qualitative discrepancies remain compared to full numerical simulations, likely because numerical summation of dipole-dipole interactions in real space is poorly convergent~\cite{Koenderink2006,Linton2010}, because  actual lattices in experiments are not 1D, and because  interactions also involve magnetic dipole-dipole coupling and magnetoelectric coupling. The minimum requirements for a simple dipole lattice model for metamaterials must necessarily include the electrodynamic  coupling between electric dipoles, magnetic dipoles as well as the cross coupling between magnetic and electric dipoles.
Here, we propose a simple model that employs exponentially convergent dipole sums and can deal with infinite 2D periodic lattices, taking any physical magnetoelectric polarizability tensor as input. The benefit of such a model is that it predicts quantitative transmission and reflection spectra that can be directly matched to data.\cite{Sersic2009,Linden2004,Rockstuhl2006,Feth2010,Decker2009,Liu2008,Decker2011}
  
This paper is organized as follows. In section~\ref{sec:theory}, we  generalize Ewald lattice sum techniques\cite{GarciadeAbajo2007}  to  point scatterers with a magnetoelectric  $6\times 6$  dynamic polarizability tensor, with interactions mediated by a 6$\times6$ Green dyadic.\cite{Sersic2011}  In section~\ref{sec:lin} we compare predicted normal incidence transmission  to measured spectra for  square and rectangular SRR lattices. In section~\ref{sec:circ} we present calculations for circular polarization at oblique incidence to evidence how single-building block pseudochirality carries over into transmission asymmetry.\cite{Sersic2012}. 

\section{Lattice theory\label{sec:theory}}
\subsection{Polarizability tensor}
We consider a 2D lattice consisting of arbitrary magnetoelectric point scatterers each described by a polarizability tensor. By definition, the polarizability relates the induced electric and magnetic
dipole moment, $\bm{p}$ and $\bm{m}$, in response to
an electric and magnetic field
$\bm{E}$ and $\bm{H}$
according to\cite{Sersic2011}
\begin{equation}
\begin{pmatrix}
  \bm{p} \\
  \bm{m}  \\
\end{pmatrix}=\alphat \begin{pmatrix}
  \bm{E} \\
  \bm{H}  \\
\end{pmatrix}.
\end{equation}
The magnetoelectric polarizability may be conveniently written as
\begin{equation}
\alphat=
\begin{pmatrix}
  \alphat_{EE} & \alphat_{EH} \\
  \alphat_{HE} & \alphat_{HH}
\end{pmatrix},
\end{equation}
where $\alphat_{EE}$ is the $3\times 3$ electric polarizability tensor that quantifies the induced electric dipole moment in response to an electric field. Similarly, $\alphat_{HH}$ describes the magnetic polarizability that quantifies the induced magnetic dipole in response to a magnetic driving field. Finally, $\alphat_{EH}$ denotes the magnetoelectric coupling that describes the induced electric dipole moment in response to a magnetic field and vice versa. We shall denote $\alphat$ the bare polarizability, since it describes the induced dipole moments in the absence of neighbouring point scatterers.
As treated in Ref. \onlinecite{Sersic2011}, $\alphat$ is subject to several constrains that we for completeness shall  briefly summarize: 
Due to reciprocity the polarizability is subject to the Onsager constraints\cite{serdyukov2001electromagnetics,Marques2002}
 \begin{equation}
   \alphat_{EE}=  \alphat_{EE}^\top,\quad
   \alphat_{HH}=  \alphat_{HH}^\top\quad
   \alphat_{EH}= -\alphat_{HE}^\top,
 \end{equation}
where the superscripted $\top$ denotes matrix transpose.
Moreover, energy conservation constrains the dynamic polarizability, in case of no Ohmic loss, to fulfill an optical theorem of the Sipe-Kranendonk\cite{Sipe1974} form
 \begin{equation}
 \frac{1}{2i}\left[\alphat-\alphat^{*\top}\right]=\frac{2}{3}k^3\alphat^{*\top}\alphat.\label{eq:optTheorem}
 \end{equation}
as derived by Belov et al.,\cite{Belov2003} and later by Sersic et al.\cite{Sersic2011}
A different way of writing this constraint is that the scalar optical theorem $\mathrm{Im}\alpha \leq 2/3 k^3 |\alpha|^2$  must hold for each eigenvalue of $\alphat$, where  equality holds in absence of loss.\cite{Sersic2011}
	Any electrostatic bare polarizability tensor $\alphat_0$, such as that derived from an Ohmically damped LC-circuit model,  may be turned into a bona fide electrodynamic polarizability that is bound by the optical theorem in \cref{eq:optTheorem} by addition of radiation damping
\begin{equation}
\alphat^{-1}=\alphat_{0}^{-1}
-\frac{2}{3}k^3 i \mathbb{I},
\end{equation}
where $k$ denotes the wavevector and $\mathbb{I}$ is the 6-dimensional identity tensor and $.^{-1}$ denotes  matrix inversion. 

\subsection{Lattice response} 
\begin{figure}
\centering
\includegraphics[width=1\columnwidth]{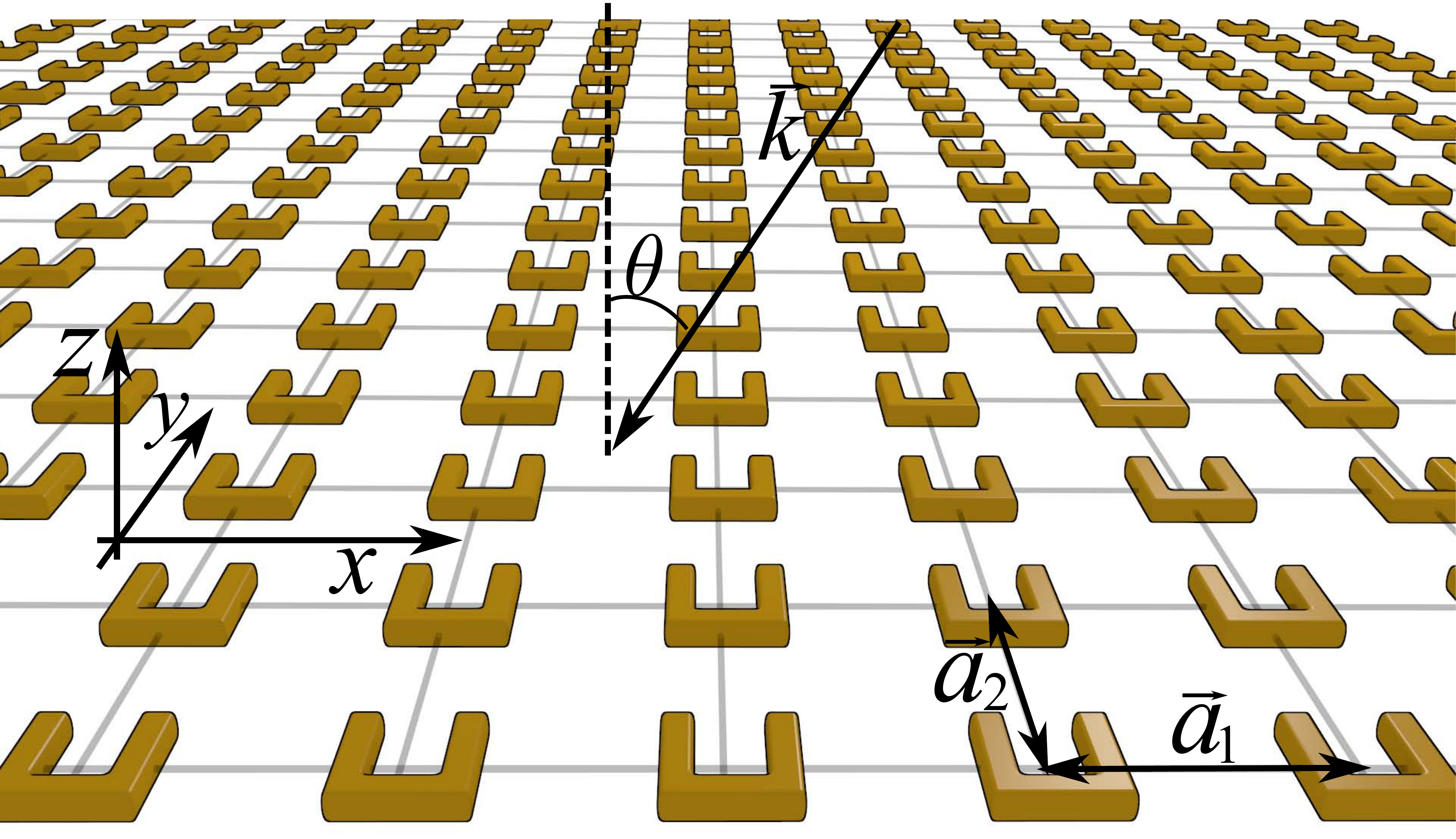}%
\caption{Illustration of the considered lattice, here sketched for split ring resonators, with a plane wave incident at an angle $\theta$.  \label{fig:setup}}%
\end{figure}

We consider the response to plane wave illumination of a 2D periodic lattice of point scatterers, which is defined by
a set of lattice vectors
$\bm{R}_{mn}=m\bm{a}_1+n\bm{a}_2$ (where $m$ and $n$
are integers, and $\bm{a}_{1,2}$ are the real space basis vectors, see \cref{fig:setup}.
The response of a particle at position $\bm{R}_{mn}$ is self-consistently set by the
incident field, plus the field of all other dipoles in the lattice according to\cite{GarciadeAbajo2007}
\begin{eqnarray}
\begin{pmatrix}
  \bm{p}_{mn} \\
  \bm{m}_{mn}  \\
\end{pmatrix}
&=&\alphat \left[
\begin{pmatrix}
  \bm{E}_{\mathrm{in}}(\bm{R}_{mn}) \\
  \bm{H}_{\mathrm{in}}(\bm{R}_{mn})  \\
\end{pmatrix} \right. \nonumber \\ & & \quad \left. + \sum_{m' \neq m, n'\neq n}
\tensor{G}^0(\bm{R}_{mn}-\bm{R}_{m'n'})
\begin{pmatrix}
  \bm{p}_{m'n'} \\
  \bm{m}_{m'n'}  \\
\end{pmatrix}
 \right] \nonumber \\
\end{eqnarray}
where $\tensor{G}^0(\bm{R}_{mn}-\bm{R}_{m'n'})$  is the $6\times 6$ dyadic Green function of the medium surrounding the split ring lattice. In this work, we take the surrounding medium to be homogeneous space.

For plane wave incidence with wave vector $\bm{k}_{||}$ and using translation invariance of the
lattice, we can substitute a Bloch wave form
$(\bm{p}_{mn},\bm{m}_{mn})^T
  =e^{i\bm{k}_{||}\cdot\bm{R}_{mn}}(\bm{p}_{00},
  \bm{m}_{00})^T$ to obtain
\begin{equation}
\begin{pmatrix}
  \bm{p}_{00} \\
  \bm{m}_{00}  \\
\end{pmatrix}
=[\alphat^{-1} -
\Gneq(\bm{k}_{||},0)]^{-1}
\begin{pmatrix} \bm{E}_{\mathrm{in}}(\bm{R}_{00}) \\
  \bm{H}_{\mathrm{in}}(\bm{R}_{00})  \\
  \end{pmatrix}
\end{equation}
Here, $\Gneq(\bm{k}_{||},0)$ is a summation of the
dyadic Green function $\tensor{G}^0$ over
all positions on the 2D periodic real space lattice barring the
origin:
\begin{equation}
\Gneq(\bm{k}_{||},\bm{r})=\sum_{m\neq 0,n\neq 0}
\tensor{G}^0({\bm{R}}_{mn}-\bm{r})e^{i\bm{k}_{||}
\cdot \bm{R}_{mn}}\label{eq:latticeSum}
\end{equation}
We will refer to the summation without exclusion of $m=n=0$ as
$\Gsum(\bm{k}_{||},\bm{r})$.  We immediately identify the factor $[\alphat^{-1} -
\Gneq(\bm{k}_{||},0)]^{-1}$ to be an \emph{effective} polarizability tensor of the \acs{SRR}, renormalized by the  lattice interactions. This is equivalent to the formulation that is didactically explained by Garc\'{i}a de Abajo in ref. \onlinecite{GarciadeAbajo2007}, however, now generalized to the magnetoelectric case.  Importantly, the summed lattice interactions not only renormalize the magnitude of  $\alpha$,  but also the relative strength of the electric and magnetic terms, and the magneto-electric cross coupling.  Since we are not aware of any reported implementation of lattice sums for the 6 $\times$ 6 dyadic Green function $\tensor{G}^0$  we supply full details in the appendix \ref{sec:appendix}. The challenging nature of the summations lies in the fact that dipole sums are poorly convergent as a real space summation due to the fact that the Green function only has a $1/r$ drop off. To overcome this, we directly follow the formulation by Linton\cite{Linton2010},  splitting the summation into a real space part and a reciprocal space part that both converge exponentially. While the work by Linton treats the Green function of the scalar Helmholtz equation\cite{Linton2010}, the necessary steps for expanding it to the $6\times 6$
dyadic Green functions are easily derived.

\subsection{Far field}
Once one has obtained the induced dipole moments, given the
incident field,  the field distribution immediately follows as\cite{GarciadeAbajo2007}
\begin{equation}
\begin{pmatrix}
\bm{E}(\bm{r})\\\bm{H} (\bm{r}) \\
\end{pmatrix}=
\begin{pmatrix}
\bm{E}_{\mathrm{in}} \\ \bm{H}_{\mathrm{in}} \\
\end{pmatrix}e^{i\bm{k}\cdot \bm{r}}
+{\cal G}(\bm{k}_{||},\bm{r})
\begin{pmatrix}
\bm{p}_{00} \\\bm{m}_{00} \\
\end{pmatrix}
\label{eq:solution}
\end{equation}
where the second term describes the scattered field. 
To find the reflected and transmitted far field amplitudes, we note that for an observation point in the far field, the Green function can be written as\cite{Novotny2008}
\begin{equation}
\tensor{G}^0(\bm{r}-\bm{R}_{mn})=k^2\frac{\exp({ik|\bm{r}-\bm{R}_{mn}|})}{|\bm{r}-\bm{R}_{mn}|} \tensor{M}_{mn}^{(\infty)}\label{eq:farfield}
\end{equation}
where $\tensor{M}_{mn}^{(\infty)}$ is a dimensionless matrix  with elements of order unity that only depends on the direction and not the length of $\bm{r}-\bm{R}_{mn}$, and which we list in appendix~\ref{app:Farfield}.
Taking the scattered field as the sum over all lattice points
\begin{equation}
\begin{pmatrix}
\bm{E}_{s}(\bm{r})\\\bm{H}_{s} (\bm{r}) \\
\end{pmatrix} = \sum_{n,m}k^2 \frac{ \exp({ik|\bm{r}-\bm{R}_{nm}|}) }{|\bm{r}-\bm{R}_{nm}|} 
e^{i\bm{k}_{||}\cdot \bm{R}_{nm}}
\tensor{M}_{nm}^{(\infty)}\begin{pmatrix} \bm{p}_{00} \\ \bm{m}_{00}\\ \end{pmatrix}\label{eq:scatField}
\end{equation}
we make the far-field expansion assuming that the orientational factor $ \tensor{M}_{mn}^{(\infty)}$ does not vary with $(n,m)$, and using the identity
\begin{equation}
\frac{\exp({ik|\bm{r}-\bm{R}_{mn}|})}{|\bm{r}-\bm{R}_{mn}|}=\frac{i}{2\pi}\int d\bm{q} 
\frac{\exp(i\bm{q}\cdot(\bm{r}_{||} -\bm{R}_{mn})  +k_z z )}{k_z}\label{eq:identity}
\end{equation}
with $k_z=\sqrt{k^2-|\bm{q}|^2}$.  Furthermore, we use the completeness relation of the lattice 
\begin{equation}\
\sum_{m,n}  e^{i\bm{k}_{||}\cdot \bm{R}_{mn}} =
 \frac{(2\pi)^2}{{\cal{A}}}\sum_{\tilde{m},\tilde{n}}
 \delta(\bm{k}_{||}-\bm{g}_{\tilde{m}\tilde{n}}).\label{eq:completeness}
\end{equation}
where ${\cal{A}}$ is the real space unit cell surface area spanned by the basis vectors $\bm{a}_1$ and $\bm{a}_2$ and $\bm{g}_{\tilde{m}\tilde{n}}=\tilde{m}\bm{b}_1+\tilde{n}\bm{b}_2$ with $\bm{b}_{1,2}$  being the reciprocal lattice basis vectors. Inserting \cref{eq:identity} and\eqref{eq:completeness} into \eqref{eq:scatField} one retrieves diffracted orders in the far field of the form
\begin{equation}
\begin{pmatrix}
\bm{E}_{s}(\bm{r})\\\bm{H}_{s} (\bm{r})\\
\end{pmatrix}
=
\sum_{\tilde{m}\tilde{n}, |\bm{k}_{\tilde{m}\tilde{n}}|\leq k}
\begin{pmatrix}
\bm{E}_{\tilde{m}\tilde{n}}\\
\bm{H}_{\tilde{m}\tilde{n}}
\end{pmatrix} e^{i\bm{k}_{\tilde{m}\tilde{n}}\cdot\bm{r}}\label{eq:farField}
\end{equation} 
where 
  $\bm{k}_{\tilde{m}\tilde{n}}=(\bm{k}_{||}+\bm{g}_{\tilde{m}\tilde{n}}, \pm \sqrt{k^2-|\bm{k}_{||}+\bm{g}_{\tilde{m}\tilde{n}}|^2})=k (\cos\phi_{\tilde{m}\tilde{n}}\sin\theta_{\tilde{m}\tilde{n}}, \sin\phi_{\tilde{m}\tilde{n}}\sin\theta_{\tilde{m}\tilde{n}},\cos\theta_{\tilde{m}\tilde{n}} )$ are the diffracted wave vectors.
The fields associated with each order are
\begin{equation}
\begin{pmatrix}
\bm{E}_{\tilde{m}\tilde{n}}\\
\bm{H}_{\tilde{m}\tilde{n}} 
\end{pmatrix}
=\frac{i 2\pi k}{A \cos\theta_{\tilde{m}\tilde{n}}} \tensor{M}(\theta_{\tilde{m}\tilde{n}},\phi_{\tilde{m}\tilde{n}}) \begin{pmatrix} 
\bm{p}_{00} \\ 
\bm{m}_{00}
\end{pmatrix}\label{eq:orderField}.
\end{equation}
Using \cref{eq:farField}, for a field incident with angles $(\theta,\phi)$ we may calculate the transmitted farfield intensity as $I_{t}= -\frac{1}{2Z_{0}}\mathrm{Re}[\bm{E}(\theta,\phi)\times\bm{H}(\theta,\phi)]\cdot \hat{z}$, where $(\bm{E},\bm{H})$ denotes the sum of the incoming $(\bm{E}\sub{in},\bm{H}\sub{in})$ and forward scattered field $(\bm{E}_{s},\bm{H}_{s})$. Similarly the reflected field is found as $I_{r}= \frac{1}{2Z_{0}}\mathrm{Re}[\bm{E}_{s}(\pi-\theta,\phi)\times\bm{H}_{s}(\pi-\theta,\phi)]\cdot\hat{z}$.
Dividing the incident intensity with the reflected/transmitted intensity we obtain the intensity reflection/transmission coefficients. For sufficiently large pitch,  grating diffraction orders will appear.  
For the common case of planar magnetic scatterers such as split rings, where the magnetic dipole moment must be perpendicular to the 2D plane, and the electric dipole must be along $x$, the  normal incidence zero-order amplitude transmission reduces to
$$
t_{xx}=1+\frac{2\pi i k }{{\cal{A}}} \left[\frac{1}{1/\alphat - \Gneq(\bm{k}_{||}=0,0)}\right]_{11}.
$$
where the subscripted $11$ denotes the first row and first column entry of the matrix, and the subscript $xx$ indicates transmission in the $x$-polarized output channel given $x$-polarized input light.

\section{Results}
\subsection{Linear polarization\label{sec:lin}}
To verify how far the simple model, presented in \ref{sec:theory}, captures the transmission properties of actual metamaterials, we compare calculated transmission spectra to measurements reported in Ref.~\onlinecite{Sersic2009} on Au split ring resonator lattices with dimensions $200\unit{nm}\times 200\unit{nm}\times30\unit{nm}$.   These split rings (made using e-beam lithography and lift-off) have a split width of 80~nm between the two arms. With the geometry illustrated in figure \ref{fig:setup}, we use a polarizability tensor of the form 
\begin{equation}
\tensor{\alpha}_{0}=\mathcal{L}(\omega)
\begin{pmatrix}
\eta_{E} & 0& \ldots & 0 & i\eta_{C}\\
0 & 0 & & & 0\\
\vdots & &\ddots & & \vdots\\
0 & & & 0 & 0\\
-i\eta_{C} & 0 & \ldots &0 & \eta_{H}
\end{pmatrix},\label{eq:barePol}
\end{equation}
where $\mathcal{L}(\omega)$ is a Lorentzian prefactor
\begin{equation}
\mathcal{L}(\omega)=V\frac{\omega_{0}^{2}}{\omega_{0}^{2}-\omega^{2}-i\omega\gamma},
\end{equation}
where $\gamma$ describes the damping rate due to Ohmic losses,  and $V$ is the physical SRR volume.  Within the chosen unit system\cite{Sersic2011},  the quantities $\eta_E, \eta_H$ and $\eta_C$ are dimensionless and directly comparable in magnitude. For implementation, we note that the polarizability tensor in Eq.~\eqref{eq:barePol} is not strictly invertible. This problem may be amended either by limiting the calculation to the $(E_{x}, H_{z})$  subspace, by employing the Moore-Penrose pseudo inverse or by substituting a small polarizability for the zeroes on the diagonal. We use the latter method.  Higher order resonances can be added to the electrostatic polarizability \emph{prior} to applying the radiation damping correction.

\begin{figure}
\centering
\includegraphics[width=1\columnwidth]{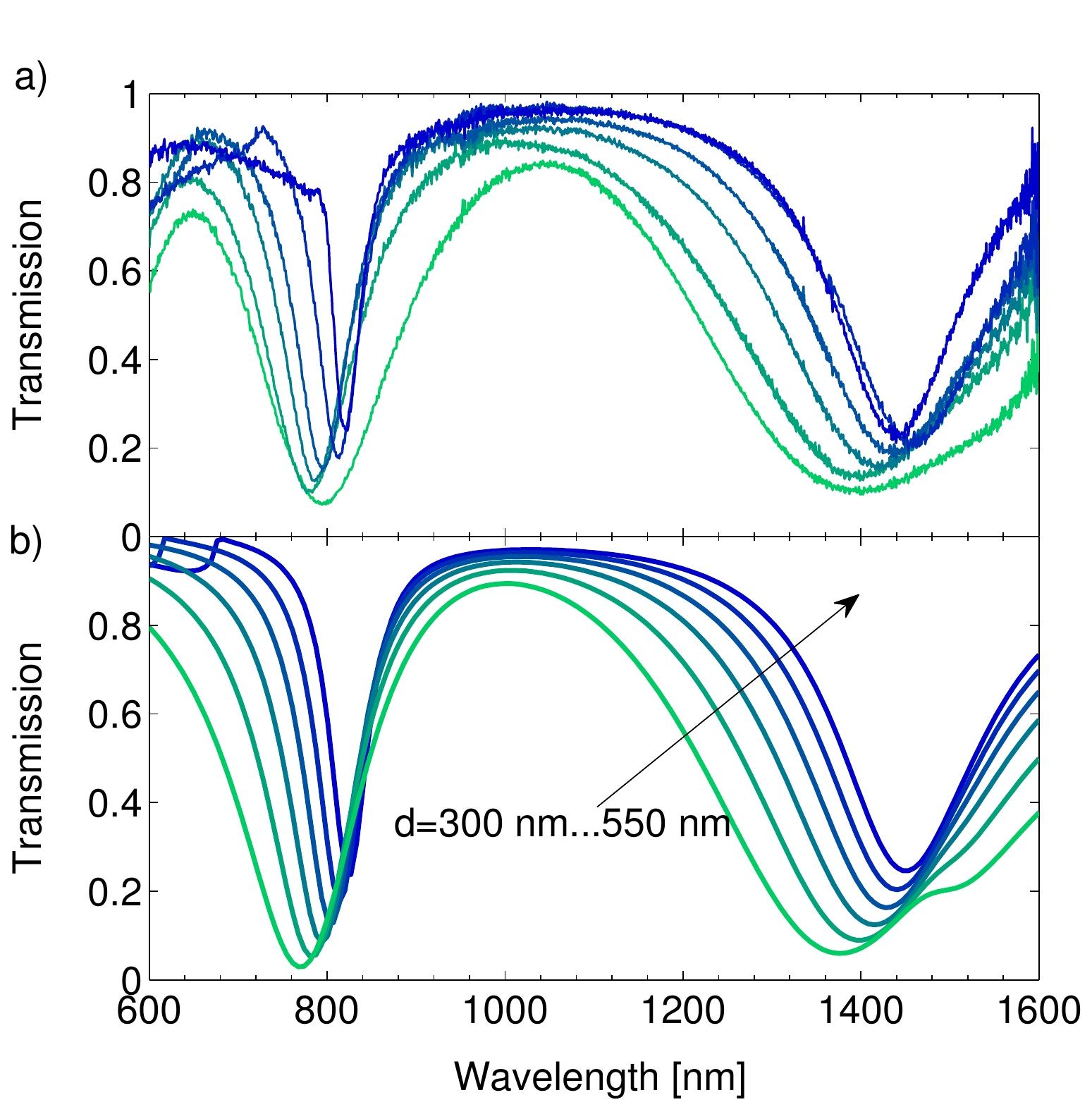}%
\caption{a) Measured transmission spectrum for a normal incidence field linearly polarized along $x$ lattice spacings between 300 nm to 550 nm in steps of 50 nm.  (see fig \ref{fig:setup} for geometry sketch).
b) Corresponding calculated transmission spectrum using $(\eta_{E},\eta_{H},\eta_{C})=(0.40, 0.26, 0.33)$ and $\lambda_{0}=1.52\unit{\mu m}$  for the magnetic resonance and $(\eta_{E},\eta_{H},\eta_{C})=(0.17, 0.00, 0.00)$ and $\lambda_{1}=0.82\unit{\mu m}$ for the plasmonic resonance using $V=200\unit{nm}\times 200\unit{nm}\times 30\unit{nm}$ and $\gamma=8.3\cdot10^{13}\unit{s^{-1}}$\label{fig:TransmissionComp}}%
\end{figure}

Figure \ref{fig:TransmissionComp}a) shows   measured normal incidence transmission versus wavelength for square lattices with pitches ranging from 300 nm to 550 nm, while \ref{fig:TransmissionComp}b) shows the corresponding calculated spectra. The data is reproduced from Ref. \onlinecite{Sersic2009}.
Two distinct resonance are observed near 1500 nm and 800 nm that are associated with respectively, the LC magnetic resonance, and a higher order plasmonic resonance, respectively\cite{Enkrich2005,Rockstuhl2006}.   For the most dilute lattices, the higher order resonance overlaps with a Rayleigh anomaly, i.e., the emergence of a grating diffraction order into the glass substrate.
Based on the lattice sum theory, presented in \ref{sec:theory}, we calculated the transmission spectrum for comparison, taking the static polarizability as the sum $\tensor{\alpha}_{a}(\omega_{a})+\tensor{\alpha}_{b}(\omega_{b})$ of two resonances (Eq.~\eqref{eq:barePol}).  For both resonances, we use a set of parameters, $\eta_{E}$, $\eta_{H}$,$\eta_{C}$ and $\omega_{0}$ common to all lattice spacings, that we obtain by  fitting all six measured spectra simultaneously by minimizing the sum of squared residuals  over the entire measured wavelength range.
In \cref{fig:TransmissionComp}b) the corresponding calculated transmission spectra are presented using fitted parameters 
$(\eta_{E},\eta_{H},\eta_{C})=(0.40, 0.26, 0.33)$ and $\lambda_{0}=1.52\unit{\mu m}$  for the  magnetic resonance and $(\eta_{E},\eta_{H},\eta_{C})=(0.17, 0.00, 0.00)$ and $\lambda_{1}=0.82\unit{\mu m}$ for the plasmonic resonance.  We discuss the confidence in these parameters further below.
Throughout this entire paper, the damping rate of gold, SRR volume and the refractive index of the surrounding medium were not fitted but fixed at $\gamma=8.3\cdot10^{13}\unit{s^{-1}}$, $V=200\unit{nm}\times200\unit{nm}\times30\unit{nm}$ and $n=1.23$.  The value $n=1.23$ reflects the average refractive index between glass and air, and is used because the lattice sum formulation as reported here can not include the actual asymmetric environment, i.e., the air-glass interface on which the split rings are situated.

 From \cref{fig:TransmissionComp} we notice that the lattice sum model   reproduces all features observed in the experimental data. Focusing on the magnetic resonance, it clearly predicts the broadening and blue shift of the resonance for decreasing pitch. From the calculated transmission we observe a second shoulder emerging for the largest density, which is only barely resolved in the experimental data. Such a resonance splitting is expected since the single SRR resonance is associated with two frequency-degenerate eigenpolarizabilities, each being a different coherent superposition of $\bm{p}$ and $\bm{m}$\cite{Sersic2011}. Increasing the density increases the magneto-electric dipole-dipole coupling between SRRs which lifts the degeneracy.   In terms of the dynamic on-resonance polarizabilities, the fitted parameters translate into $|\alpha_E|=3.8V$, $|\alpha_H|=2.5V$ and $|\alpha_C|=3.2V$. The extracted parameters indicate that the LC resonance is primarily electric in nature, and that the bi-anisotropy  $\eta_C$ makes it significantly easier for the electric field to induce a magnetic dipole than it is for the magnetic field.  

To quantify the agreement between our new  calculation methods and previously reported measurement data, we extracted the center frequency, the resonance linewidth and the extinction cross section of the magnetic resonance on three types of sample sets: one with a square grid, where both $d_{x}$ and $d_{y}$ were changed equally over each sample, one with a rectangular grid with $d_{x}=500\unit{nm}$ and $d_{y}$ varying and similarly one with a rectangular grid where $d_{y}=500\unit{nm}$ and $d_{x}$ varying. In order to correct for  the well-known electron-beam lithgography artefact that object density affects the required dose for realizing a specific feature, i.e., the so-called proximity effect, we fabricated samples at different e-beam dose factors, and used image analysis software to select arrays in which SRRs had identical dimensions (arm length, base length, gap width, gap depth)  to within better than 5 nm.  As reported in Ref.~\onlinecite{Sersic2009},  the gap between the arms for this set of samples is significantly larger at 100 nm, than it is for ones presented in \cref{fig:TransmissionComp}. The center frequency, resonance linewidth and effective extinction cross section were extracted from the data by fitting a Lorentzian to the transmission resonance. The effective extinction cross section per split ring is defined as $\sigma\sub{ext}=(1-T\sub{min})d^{2}$, with $T\sub{min}$ being the measured value of transmission at the transmission minimum.
\begin{figure}
\centering
\includegraphics[width=1\columnwidth]{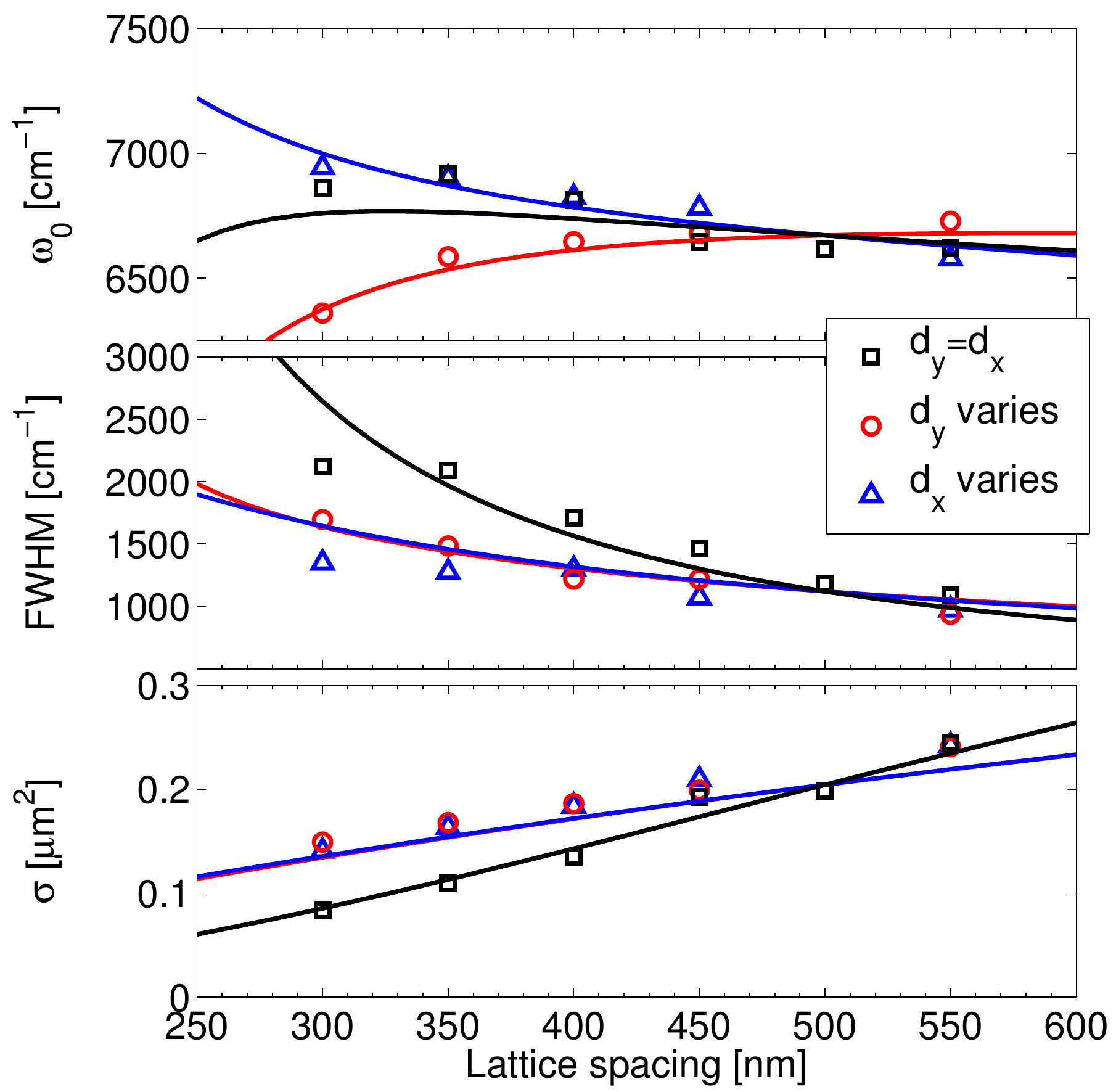}%
\caption{(Color online) Comparison of calculated (lines) and measured (markers) center frequency (a), resonance linewidth (b) and extinction cross section per split ring (c). Corresponding fit parameters were $(\eta_{E},\eta_{H},\eta_{C})=(0.63\pm0.01, 0.12\pm 0.02, 0.28\pm0.03)$ and $\lambda_{0}=1.57\pm 0.003\unit{\mu m}$ setting $V=200\unit{nm}\times 200\unit{nm}\times 30\unit{nm}$ and $\gamma=8.3\cdot10^{13}\unit{s^{-1}}$. In all plots the line color and symbol shape indicate square lattices (square symbol, black lines), resp. lattices with $d_x$ fixed to 500 nm and $d_y$ varying (circular symbol, red), and vice versa (triangular symbol).
 \label{fig:crossWidthCenter}}%
\end{figure}
To evaluate the theory,  we follow a procedure identical to the one followed for \cref{fig:TransmissionComp}. In particular, the parameters $\eta_E, \eta_H, \eta_C$ and $\omega_0$ were obtained by simultaneously fitting all measured spectra of both square and rectangular lattices, by minimizing the sum of the summed squared residual of each transmission spectrum. Subsequent to fitting the spectra, the center frequency, resonance linewidth and extinction cross section were extracted from the calculated transmission spectra exactly as done for the experimental measurements.
\Cref{fig:crossWidthCenter}  shows the density dependence of the transmission resonance frequency, linewidth and effective SRR extinction cross section, respectively, as predicted by the lattice sum model together with the values extracted from experiment. The lattice sum calculation  qualitatively reproduces the blueshift (redshift) when varying $d_{x}$ ($d_{y}$), while for the square lattice we observe some discrepancy for the shortest lattice constants. We attribute this discrepancy to the fact that the shortest pitch square lattice sample is the densest, with spacing between split rings approximately half their diameter. From estimates for coupled plasmon particles,  at and below this spacing the dipole approximation breaks down\cite{Park2004}. Considering the resonance linewidth in figure \ref{fig:crossWidthCenter}b) we first note that since the Ohmic damping does not depend on the coupling in an electrostatic model, the FWHM broadening with decreasing lattice spacing can only be explained by the radiation damping in an electrodynamic picture, which the lattice sum model fully takes into account and is in excellent agreement with the measurements. Finally, the trend of a marked increase of effective cross section with reduced density is evident, with excellent agreement between theory and measurement. 
The meaning of the strong dependence of th effective cross section per split ring on pitch is that the effective cross section is bounded from above by the single split ring extinction cross section ($\sim 0.3 \unit{\mu m}^{2}$, see ref. \onlinecite{Husnik2008}) for dilute lattices, and  by the unit cell area for dense lattices. As the lattice is made denser,  the unit cell area becomes smaller than the single object cross section.  As the unit cell area is further decreased, superradiant damping sets in that increases the FWHM and at the same time diminishes the effective extinction per split ring to be essentially pinned at the unit cell area.

We note that the theoretical values of the center frequency, linewidth and extinction cross section were extracted by fitting full spectra, i.e., by performing a nonlinear least squares fit  to match measured and calculated frequency dependent transmission $T(\omega)$.  An alternative fit procedure would be to not base the fit merit function on the deviation in $T(\omega)$,  but rather to only fit center frequency, width and extinction cross section as extracted to data to those extracted from calculated spectra.  On basis of the fit, we conclude that the parameters that best describe the experiment are $(\eta_{E},\eta_{H},\eta_{C})=(0.63\pm0.01, 0.12\pm 0.02, 0.28\pm0.03)$ and $\lambda_{0}=1.57\pm 0.003\unit{\mu m}$,  where the stated accuracies are the $95\%$ confidence interval.  The parameters are somewhat different to those obtained from the experiment in Fig.~\ref{fig:TransmissionComp}, and correspond to on-resonance dynamic polarizabilities of $|\alpha_E|=4.5V$, $|\alpha_H|=0.82 V$ and $|\alpha_C|=2.0V$. We attribute the larger ratio between electric and magnetic response to the larger split width. We found that relaxing the constraint in the fit to the requirement that only the three extracted parameters, and not necessarily the entire spectrum be fitted optimally in the least squares sense, does not yield a substantially improved value for the  $\eta$-parameters. Ultimately, the  reliability of the parameters is limited by our treatment of the dielectric environment. While the environment is in fact asymmetric (air-glass interface), we take the environment to be homogeneous with index equal to the average of both media.

\subsection{Circular polarization\label{sec:circ}}
As discussed in ref. \onlinecite{Sersic2011},   the single SRR  eigenpolarizabilities and eigenvectors of the polarizability tensor have special significance.  In particular, the eigenpolarizabilities point to a largest, and a smallest exctinction cross section that can be addressed if the illumination field is chosen to equal the correct coherent mixture of $E_x$ and $H_z$ that the eigenvector prescribes.  When $|\eta_{C}|>0$  the eigenvectors of the polarizability tensor correspond to oblique incidence circularly polarized light, implying a handed response in scattering.  The existence of `bi-anisotropy' , \emph{i.e.} $\eta_C\neq 0$, was known from the outset in the field of metamaterials\cite{Katsarakis2004}.  As predicted in ref. \onlinecite{Sersic2011}, and realized experimentally in ref. \onlinecite{Sersic2012} the \emph{strength} of this effect can be directly probed using circular polarized light. Here we present lattice sum transmission calculations using circular polarized incident light that confirm a strongly handed extinction under oblique incidence\cite{Sersic2012}. This comparison has no adjustable parameter since we take as input the polarizability tensor retrieved from the normal incidence density dependent data discussed above.

\begin{figure}
\centering
\includegraphics[width=1\columnwidth]{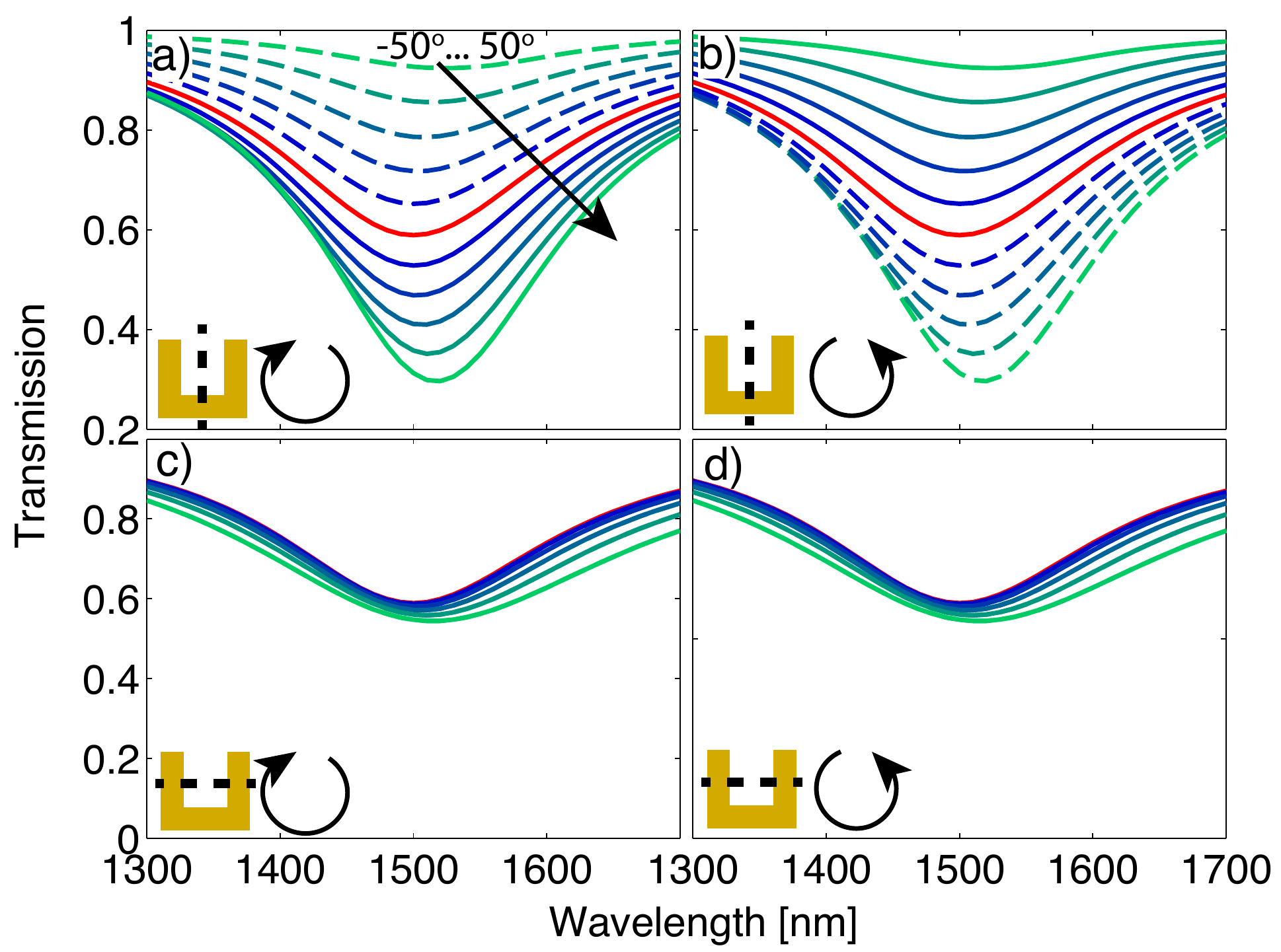}%
\caption{(Color online) Calculated transmission versus wavelength for circularly polarized light incident at angles from $-50^{\circ}$ to $50^{\circ}$ in steps of $10^{\circ}$ for a square lattice with $d=500\unit{nm}$. Negative (positive) angles are plotted as dashed (solid) lines and normal incidence is marked as solid red. a) (b)) Right (left) hand polarized light with incident angle with rotational axis along the symmetry axis of the SSR, as depicted in the inset. c) (d)) Right (left) hand polarized light with incident angle with rotational axis perpendicular to the symmetry axis of the SRR, as depicted in the inset.\label{fig:TransAngle}}%
\end{figure}

Figure \ref{fig:TransAngle} shows the calculated transmission spectrum for various incident angles using the same parameters as for \cref{fig:TransmissionComp}.
Firstly, we note that the transmission spectra, when changing the incident angles around the symmetry axis of the SRR (\cref{fig:TransAngle}a)-b)), reveal a strong angular dependence, while incident angles perpendicular to the symmetry axis reveal only a weak angular dependence.   The strong angular dependence is strongly asymmetric around normal incidence, with transmission going from barely suppressed to very strong extinction when going form negative to positive angles for righthanded polarization (reversed behavior for opposite handedness).
This is consistent with experimental results in Ref.~\onlinecite{Sersic2012}.  From a LC-circuit point of view, at oblique incidence angles the split ring is driven both by $E_{x}$ and $H_{z}$,  and the handedness determine whether the phase difference between the $E_{x}$ and the $H_{z}$ field  is such that the two driving terms for the capacitor and the current loop add up, or cancel. For rotations perpendicular to the symmetry axis [\cref{fig:TransAngle}c)-d)], no such phase difference is present.  On basis of group theory arguments, it was first noted by \textcite{Verbiest1996} that  indeed a two-dimensional lattice can show optical activity in spite of the building block being achiral. This has later been referred to as an extrinsic optical activity by~\textcite{Plum2009} and pseudochirality by \textcite{Tretyakov1998} and \textcite{Sersic2012}.

It has been argued by \textcite{Gompf2011} that for lattices with $d\sim\lambda$   spatial dispersion, an effect that is fully contained in our lattice sum approach, may conspire to induce handedness in the transmission, regardless of the shape of the building block of the lattice.  However, with the disappearance of optical activity for rotations perpendicular to the symmetry axis, seen in figure \ref{fig:TransAngle}(c-d), we conclude that it is indeed the building block that causes the handed behavior.  The fact that the contrast in transmission is large is due to the fact that one of the two eigenvalues of the split ring polarizabilities vanishes at the given large cross coupling.

\begin{figure}
\centering
\includegraphics[width=1\columnwidth]{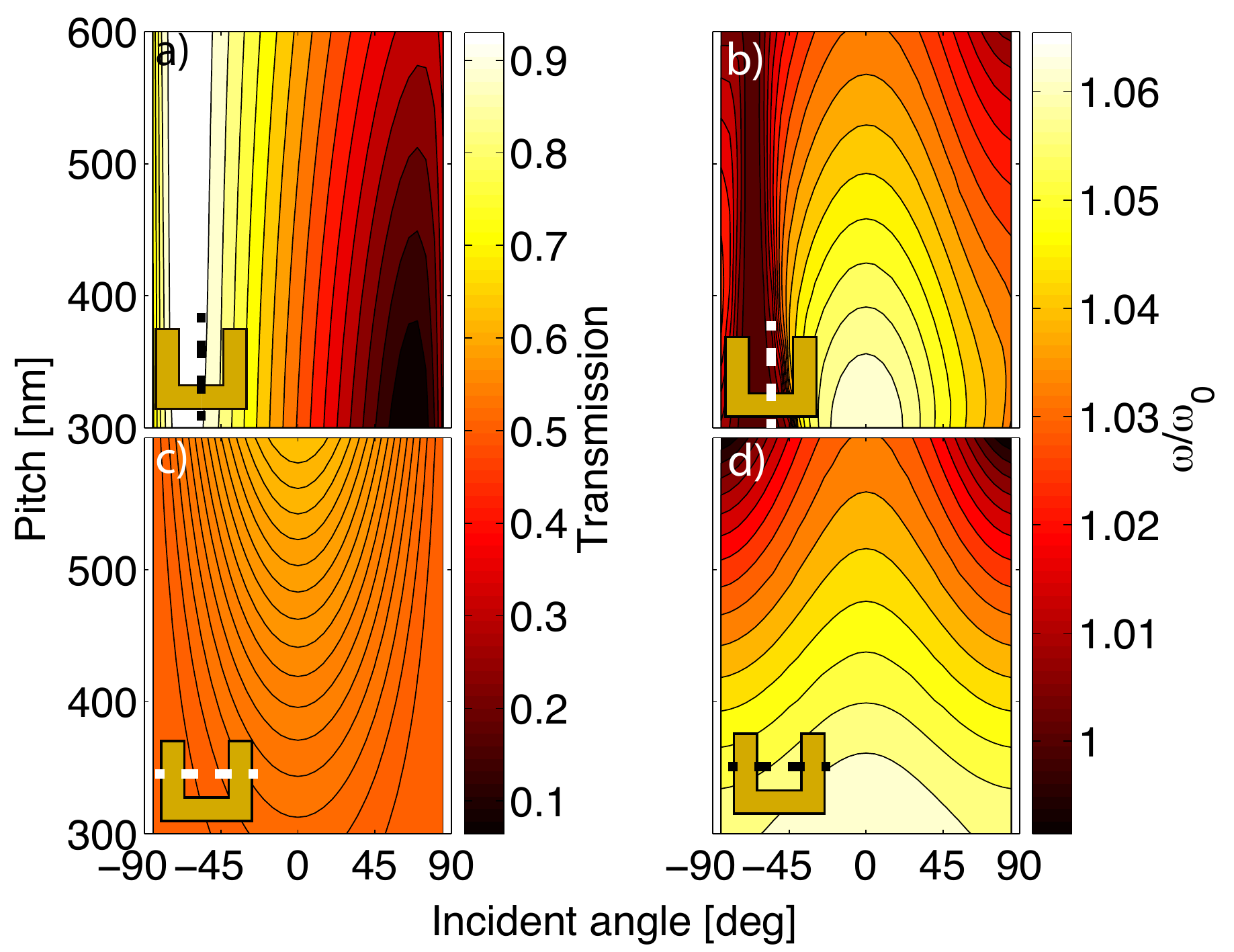}%
\caption{(Color online) Left column: Calculated minimum transmission, $T\sub{min}$, of a square lattice as a function of incident angle and lattice pitch, for an incident right hand polarized field. Right column: The associated resonance frequency normalized with the single SRR resonance frequency. Top (bottom) row: incident angle with rotational axis parallel (perpendicular) the symmetry axis of the SSR, as depicted in the inset. \label{fig:extCross}}%
\end{figure}

In \cref{fig:extCross}(a,b), the calculated minimum transmission, $T\sub{min}$, is plotted as a function of incident angle and lattice pitch, for  right-handed input polarization and incident angle with rotational axis parallel (perpendicular) the symmetry axis of the SSR. 
Considering \cref{fig:extCross}a) we first note that for any given lattice spacing, the deepest transmission minimum is reached at strongly positive angle (above 60$^\circ$), while the lattice is almost transparent (90\% transmission or more) at sharply negative angles.
  The fact that the angle of maximum and minimum transmission is rather insensitive to the lattice pitch indicates that while dipole-dipole interactions in the lattice may change the resonance frequency, width and strength,  they do not strongly modify the angle for addressing the highest pseudochiral contrast.
Comparing the resonance frequency shift in  \cref{fig:extCross}b) with the value of the transmission on resonance
$T\sub{min}$ in \cref{fig:extCross}a) it is seen that for angles close to the point where the lattice is almost transparent,  the dipole-dipole coupling induced frequency shift vanishes. This realization is consistent with the fact that at incident fields near transparency, hardly any dipole moment is set up.  Conversely we note that  for any given lattice pitch, the maximum frequency shift is located at incidence normal to the lattice, and not at the angle where the transmission minimum is deepest. This conclusion is not easily explained in a simple dipole hybridization model\cite{Prodan2003}, since the frequency shift is a complex interplay of partially cancelling transverse and longitudinal electric dipole coupling (along $\hat{\bm{x}}$ and $\bm{\hat{y}}$ respectively),  a weaker  transverse magnetic dipole coupling (along $\hat{\bm{z}}$, as well as magnetoelectric coupling  that depends on the relative phase with which $\bm{p}$ and $\bm{m}$ are driven.

\begin{figure}
\centering
\includegraphics[width=1\columnwidth]{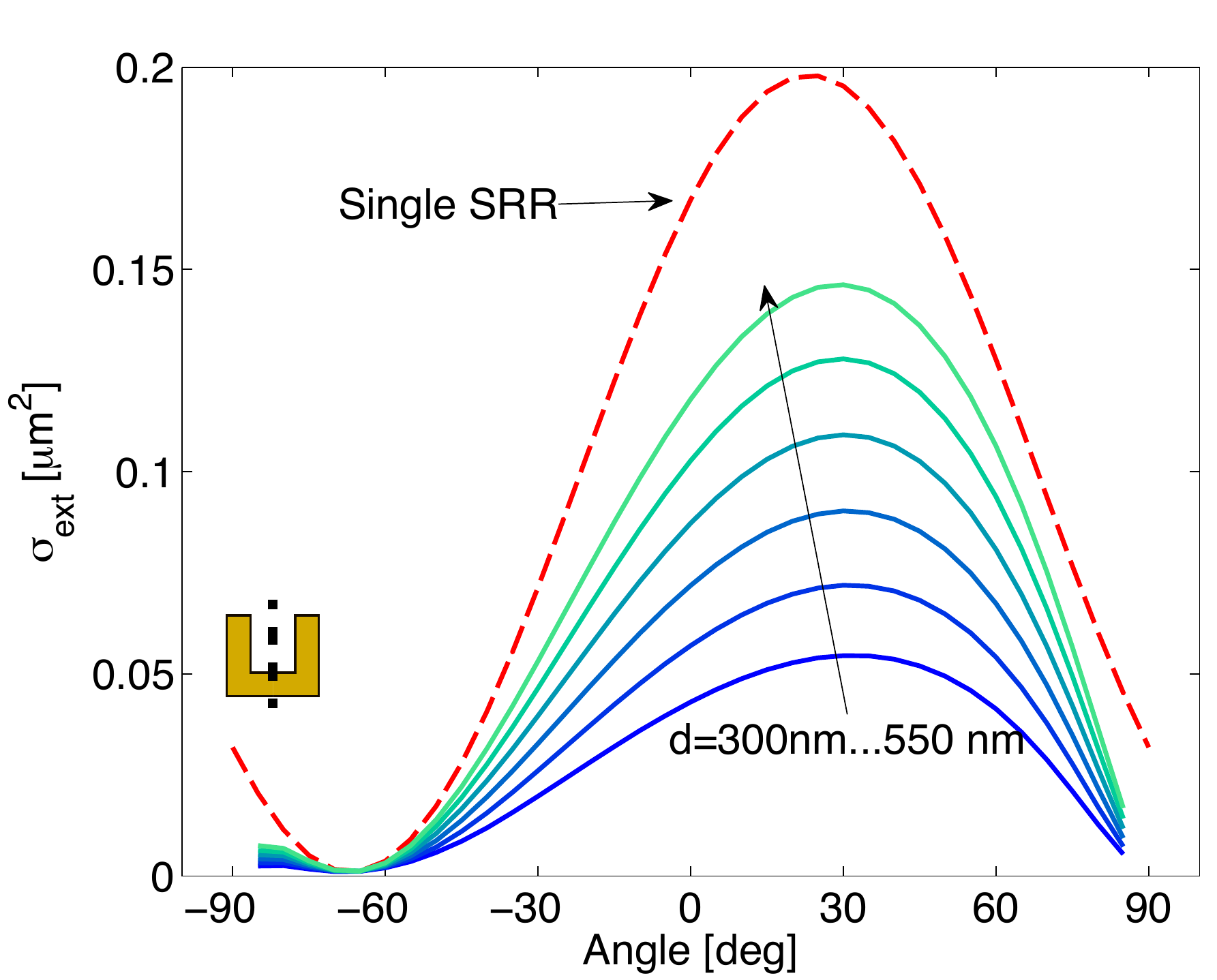}%
\caption{(Color online) Calculated extinction cross section of a square lattice as a function of incident angle and lattice pitch for various lattice spacings $d$. Rotation is around the the symmetry axis of the SRR. Dashed and solid lines are for the single SRR  and full lattice, respectively.\label{fig:extCrossComparison}}%
\end{figure}

In order to compare the full lattice calculation with those of a single SRR we calculated the extinction cross section as a function of input incident angle for six lattice spacings for the two scenarios. For the single SRR we calculated the extinctions cross section from the work done by the incident field, $W=2\pi k\{\mathrm{Re}(\bm{E}\sub{in},\bm{H}\sub{in})\cdot\mathrm{Im}[(\bm{\alpha}(\bm{E}\sub{in},\bm{H}\sub{in})^{\top}]  -  \mathrm{Im}(\bm{E}\sub{in},\bm{H}\sub{in})\cdot\mathrm{Re}[\bm{\alpha}(\bm{E}\sub{in},\bm{H}\sub{in})^{\top}]\}/Z$ divided with the input intensity, $|\bm{E}\sub{in}|^{2}/(2Z)$, where $Z$ is the impedance of the surrounding material. For the lattice calculations, we define the effective extinction cross section per SRR as   $\sigma\sub{ext}=(1-T\sub{min})d_{x}d_{y}\cos\theta$.   We note that the factor $\cos\theta$ in this definition needs to be included to account for  the simple geometrical projection argument that at larger angle an incident  beam of the same diameter intersects a larger set of split rings. The extracted effective extinction per split ring is presented in \cref{fig:extCrossComparison} for the case of rotation around the the symmetry axis of the SRR.
For the single SRR, the angular dependence on the extinction cross section can be characterized by a cosine shifted by roughly $20^{\circ}$ from the sample normal.\cite{Sersic2012}    This angle is much smaller than the angle away from the sample normal at which the maximum and minimum transmission is reached.  This is only an apparent contradiction, since  the trivial $\cos\theta$  projection effect  causes a substantial additional skewing of the angular asymmetry in  \emph{transmission} in comparison to the asymmetry in per building-block extinction.  Indeed, the effective extinction cross section per split ring, \emph{corrected} for the $\cos\theta$ projection factor, closely resembles the single SRR angle-dependent extinction,  apart from being increasingly suppressed in amplitude for decreasing lattice pitch. This suppression of the peak extinction with increasing density is a  consequence of superradiant damping exactly as also evident for the normal incidence data in Fig.~\ref{fig:crossWidthCenter}c).  Even for the largest lattice spacings there remains a significant difference between the extinction cross section of a single SRR and a SRR lattice, pointing to the importance of renormalization of the split ring response by retarded coherent interactions  in full lattice calculations  even when considering a dilute 2D metamaterial.

\section{Conclusion}
We have presented calculations of the full electromagnetic response of an infinite 2D magneto-electric dipole lattice with consistent treatment of radiation damping, retardation and energy conservation. The model was compared with recently published transmission data of a \acf{SRR} lattices with different lattice spacings. The model accounts excellently for the density dependent collective resonance frequency,  spectral width, and effective extinction cross section per split ring, in addition to capturing the strong pseudochiral response that fingerprints bi-anisotropic cross coupling.  The model that we presented can be easily extended to deal with  diffractive effects that occur at larger pitch,   the emergence of surface lattice resonances\cite{Rodriguez2011,Lozano2013}, and to stacks of lattices or  metasurfaces with more than one element per unit cell. In particular, we anticipate that the model is a semi-analytical tool  to explore the emergence, spectral and spatial dispersion of  $\bm{\epsilon}$  and $\bm{\mu}$  and their dependence on the  density and thickness of 3D metamaterials in a fully self-consistent electrodynamic multiple scattering approach.

\begin{acknowledgments}
This work is part of the research program of the “Stichting
voor Fundamenteel Onderzoek der Materie (FOM)”, which
is financially supported by the “Nederlandse Organisatie
voor Wetenschappelijk Onderzoek (NWO)”. A.F.K. gratefully acknowledges a NWO-VIDI grant for financial support. P.L. acknowledges support by the Carlsberg Foundation as well as the Danish Research Council for Independent Research (Grant No. FTP 11-116740).
\end{acknowledgments}

\appendix

\section{Sums of magneto-electric Dyadic Greens function}\label{sec:appendix}

The sum presented in Eq.~\eqref{eq:latticeSum}, requires special attention since it converges poorly. The problem has been treated extensively in \onlinecite{Linton2010} and utilizes a technique pioneered by Ewald. The technique consists in splitting a poorly convergent sum into two convergent terms, $\Gsum^{(1)}$ and $\Gsum^{(2)}$, which are exponentially convergent. Specifically, considering the sum
\begin{equation}
\Gamma(\bm{k}_{||},\bm{r})=\sum_{m,n}
G^0({\bm{R}}_{mn}-\bm{r})e^{i\bm{k}_{||}
\cdot \bm{R}_{mn}}\label{eq:scalLatticeSum}
\end{equation}
where the scalar Green function is
\begin{equation}
G^0({\bm{R}}_{mn}-\bm{r})=\frac{e^{ik
|\bm{R}_{mn}-\bm{r}|}}{|\bm{R}_{mn}-\bm{r}|}.
\end{equation}
we may rewrite this as
\begin{equation} \sum_{m,n}
\frac{e^{ik
|\bm{R}_{mn}-\bm{r}|}}{|\bm{R}_{mn}-\bm{r}|}
e^{i\bm{k}_{||}\cdot\bm{R}_{mn}}=\Gamma^{(1)}+\Gamma^{(2)}.
\end{equation}
Here
\begin{subequations}
\begin{multline}
\Gamma^{(1)}=\frac{\pi}{{\cal{A}}}\sum_{\tilde{m}\tilde{n}}\left\{
\frac{e^{i(\bm{k}_{||}+g_{\tilde{m}\tilde{n}})\cdot
\bm{r}_{||}}}{k^z_{\tilde{m}\tilde{n}}}\right.\\
\cdot\left[
e^{ik^z_{\tilde{m}\tilde{n}}|z|}\mathrm{erfc}\left(\frac{k^z_{\tilde{m}\tilde{n}}}{2\eta}+|z|\eta\right)\right.\\
+
\left.\left. e^{-ik^z_{\tilde{m}\tilde{n}}|z|}\mathrm{erfc}\left(\frac{k^z_{\tilde{m}\tilde{n}}}{2\eta}-|z|\eta\right) \right]\right\}
\label{eq:cylindricalsum}
\end{multline}
and  \ 
\begin{multline}
\Gamma^{(2)}=
\sum_{mn}\left\{\frac{e^{i\bm{k}_{||}\cdot\bm{R}_{mn}}}{2\rho_{mn}} 
\cdot\left[ e^{ik\rho_{mn}}\mathrm{erfc}\left(\rho_{mn}\eta
+\frac{ik}{2\eta}\right) \right.\right.\\
+ \left.\left. e^{-ik\rho_{mn}}\mathrm{erfc}\left(\rho_{mn}\eta
-\frac{ik}{2\eta} \right) \right]\right\},
\label{eq:sphericalsum}
\end{multline} 
\end{subequations}
where we used $\bm{r}=(\bm{r}_{||},z)$,
$k=\omega/c$, $k^z_{\tilde{m}\tilde{n}}=
\sqrt{k^2-|\bm{k}_{||}+\tensor{g}_{\tilde{m}\tilde{n}}|^2}$, and
$\rho_{mn}=|\bm{R}_{mn}-\bm{r}_{||}|$. Convergence of Eq.~\eqref{eq:sphericalsum} and Eq.~\eqref{eq:cylindricalsum} follows from the asymptotic expansion of the  error function revealing $z\,\mathrm{erfc}(z)\sim \exp(-z^{2})$ for $z\rightarrow\infty$.\cite{Linton2010}
The parameter $\eta$ can be chosen for optimal convergence, and
should be set around $\eta=\sqrt{\pi}/{a}$, where $a$ is the lattice
constant. Naturally, the cut off  for the summation over $m$ and $n$ must be chosen at least bigger than the number of propagating grating diffraction orders one expects.For our calculations on metamaterials, with essentially no grating orders, i.e., $ka\leq 2\pi$, we already obtained converged lattice sums for  $|m,n|\leq 5$.

The dyadic lattice sums in Eq.~\eqref{eq:latticeSum} are easily generated by noting that the
scalar Green function 
\begin{equation}
G(\bm{r},\bm{r'})=\frac{\exp\left({ik|\bm{r}-\bm{r}'|}\right)}{|\bm{r}-\bm{r}'|}
\end{equation}
 sets the dyadic
Green function via
\begin{equation}
\tensor{G}^0(\bm{r}-\bm{r}') =
\begin{pmatrix}
\mathbb{I} k^2 +\nabla\otimes\nabla &
-ik\nabla \times  \\
 ik\nabla \times   &  \mathbb{I} k^2
+\nabla\otimes\nabla \\
\end{pmatrix} G(\bm{r},\bm{r}')
\label{eq:derive}
\end{equation}
where $\mathbb{I}$ indicates the $3\times 3$ identity matrix and $\otimes$ denotes the outer product.
The derivatives can be simply pulled into each exponentially
convergent sum to be applied to each term separately, and are most
easily implemented in practice by noting that the sum
$\Gamma^{(2)}$ only depends on radius in spherical coordinates
$\rho_{mn}$, while the sum in $\Gamma^{(1)}$ only depends on radius and height in cylindrical coordinates. For these coordinate systems the differential operator in Eq.~(\ref{eq:derive}) take particularly simple forms. For spherical coordinates this form reads
\begin{subequations}
\begin{multline} (\mathbb{I} k^2 +\nabla\nabla )F(r) =  
\mathbb{I}\left[k^2  F(r) + \frac{1}{r}\frac{d}{dr} F(r)\right]\\
+
\begin{pmatrix}
x^2 & xy & xz \\
xy & y^2 & yz \\
xz & yz  & z^2 \\
 \end{pmatrix}
 \frac{1}{r}\frac{d}{dr}( \frac{1}{r}\frac{d}{dr} F(r))
 \end{multline}
 and
 \begin{multline}
 -ik\nabla \times F(r)  =   ik \begin{pmatrix}
0 & z & -y    \\
-z & 0 & x \\
y & -x  & 0 \\
 \end{pmatrix}\frac{1}{r}\frac{d}{dr} F(r),
\end{multline}
\end{subequations}
which can be directly applied to the summands in
Eq.~(\ref{eq:sphericalsum}).
 For cylindrical coordinates the differential  form reads
 \begin{subequations}
\begin{multline} 
(\mathbb{I} k^2 +\nabla\otimes\nabla )
e^{i\bm{k}\cdot\rho} g(z)=\\
\begin{pmatrix}
k^2-k_x^2 & -k_xk_y & 0 \\
 -k_xk_y & k^2-k_y^2 & 0 \\
0 &0  & k^2 \\
 \end{pmatrix}e^{i\bm{k}\cdot\bm{r}_{||}} g(z) \\
+\begin{pmatrix}
0 & 0 & ik_x  \\
 0 & 0 & ik_y \\
ik_x &ik_y  & 0  \\
 \end{pmatrix}e^{i\bm{k}\cdot\bm{r}_{||}} \frac{\mathrm{d} g(z)}{\mathrm{d}z}\\
+\begin{pmatrix}
0 & 0 & 0  \\
 0 & 0 & 0 \\
0 & 0  & 1  \\
\end{pmatrix}e^{i\bm{k}\cdot\bm{r}_{||}} \frac{\mathrm{d}^2 g(z)}{\mathrm{d}z^2}
\end{multline}
and
\begin{multline}
  -ik\nabla \times e^{i\bm{k}\cdot\bm{r}_{||}} g(z)  =\\
 \begin{pmatrix}
0 & 0 & -k k_y  \\
 0 & 0 &  kk_x \\
kk_y &-kk_x  & 0  \\
\end{pmatrix}
e^{i\bm{k}\cdot\bm{r}_{||}} g(z) \\
+
\begin{pmatrix}
0 & ik & 0  \\
 -ik & 0 & 0 \\
0 & 0  & 0  \\
\end{pmatrix}e^{i\bm{k}\bm{r}_{||}} \frac{d g(z)}{dz}
\label{eq:cyldiff}\end{multline}
 \end{subequations}
which can be directly applied to evaluate the dyadic equivalent of
Eq.~(\ref{eq:cylindricalsum}).

\section{Far field}\label{app:Farfield}
Carrying out the differentiation in Eq.~\eqref{eq:derive} keeping only terms with $(|\bm{r}-\bm{r}'| k)^{-1}$ we get
\begin{equation}
\tensor{G}^0_{\infty}(\bm{r}-\bm{R}_{mn}) =\tensor{M}_{mn}^{(\infty)} \,G(\bm{r},\bm{R}_{mn})
\end{equation}
where (with $\bm{\xi}=(\bm{r}-\bm{R}_{mn})/|\bm{r}-\bm{R}_{mn}|$ )
\begin{subequations}
\begin{align}
\tensor{M}_{mn}^{(\infty)}&=
\begin{pmatrix}
\tensor{A} &  \tensor{B}\\
-\tensor{B} & \tensor{A}
\end{pmatrix}
\qquad\textrm{with}\\
\tensor{A}&=\mathbb{I}-\bm{\xi}\otimes\bm{\xi},\quad\textrm{and} \quad
 \tensor{B}&=
\begin{pmatrix}
0 & \xi_{z} &-\xi_{y} \\
-\xi_{z} & 0 & \xi_{x}\\
\xi_{y} &-\xi_{x} & 0
\end{pmatrix},
\end{align}
\end{subequations}

\begin{acronym}
\acro{FDTD}{finite difference time domain}
\acro{SRR}{split ring resonator}
\end{acronym}

\bibliography{AMOLF-LatticesumPaper}

\end{document}